\documentclass[10pt,a4paper,reqno]{amsart}
\usepackage{acronym}
\usepackage{amsfonts}
\usepackage{amsmath}
\usepackage{amssymb}
\usepackage{amsthm}
\usepackage{bm}
\usepackage{braket}
\usepackage{cite}
\usepackage{color}
\usepackage{geometry}
\usepackage{graphicx}
\usepackage{hyperref}
\usepackage{mathrsfs}
\usepackage{mathtools}
\usepackage{setspace}
\usepackage{stmaryrd}
\usepackage{subcaption}
\usepackage{tikz}
\usepackage{CJKutf8} 
\usepackage{epsfig}
\usepackage{setspace}
\usepackage{booktabs}
\usepackage{threeparttable}
\usepackage{diagbox}
\usepackage{float}

\newgeometry{top=2cm,bottom=2cm,outer=1.5cm,inner=1.5cm}
\newcommand{\bse}{\begin{subequations}}
\newcommand{\ese}{\end{subequations}}

\numberwithin{equation}{section}

\title[Physics-informed
neural networks method in high-dimensional integrable systems]{Physics-informed
neural networks method in high-dimensional integrable systems}
\author{Zhengwu Miao,}
\address[MZ]{School of Mathematical Sciences, Shanghai Key Laboratory of Pure Mathematics and Mathematical Practice, and Shanghai Key Laboratory of Trustworthy Computing \\
East China Normal University \\ Shanghai 200241 \\ People's Republic of China}
\author{Yong Chen$^*$}
\address[YC]{School of Mathematical Sciences, Shanghai Key Laboratory of Pure Mathematics and Mathematical Practice, and Shanghai Key Laboratory of Trustworthy Computing \\
East China Normal University \\ Shanghai 200241 \\ People's Republic of China}
\address[YC]{College of Mathematics and Systems Science \\ Shandong University of Science and Technology \\ Qingdao 266590 \\ People's Republic of China}
\email{ychen@sei.ecnu.edu.cn}

\begin{document}

\begin{abstract}
In this paper, the physics-informed neural networks (PINN) is applied to high-dimensional system to solve the $(N+1)$-dimensional initial boundary value problem with $2N+1$ hyperplane boundaries. This method is used to solve the most classic $(2+1)$-dimensional integrable Kadomtsev-Petviashvili (KP) equation and $(3+1)$-dimensional reduced KP equation. The dynamics of $(2+1)$-dimensional local waves such as solitons, breathers, lump and resonance rogue are reproduced. Numerical results display that the magnitude of the error is much smaller than the wave height itself, so it is considered that the classical solutions in these integrable systems are well obtained based on the data-driven mechanism.
\end{abstract}

\maketitle

\section{Introduction}

Nonlinear problems play an important role in mathematical physics and engineering fields such as fluid mechanics, plasma physics, and optical fiber communication\cite{TKRJ15,DA78,AA14,AD98}. The nonlinear phenomenon in the natural world are abstracted as simple mathematical and physical models, which are expressed as nonlinear partial differential equations (NPDEs). In recent decades, the solution of NPDEs has been the focus of researchers\cite{VM91,MP91,CW02,H04}.

With the explosive growth of data resources, deep learning technologies with the advantage of $big\ data$ has been widely used in various fields including image recognition\cite{AIG12}, cognitive science\cite{BRJ15}, natural language processing\cite{YYG15}, etc., and has made revolutionary progress. Convolutional neural networks (CNN), recurrent neural networks (RNN) and other network structures that have been continuously developed are more inclined to solve specific learning tasks\cite{AIG12,WIO1409}. However, the high cost of data acquisition makes traditional deep learning algorithms face huge challenges in complex physics and engineering analysis. Recently, the deep learning framework based on physical constraints (physics-informed neural networks) proposed by M. Raissi et al. provides a new idea for the learning of $small\ data$ regime\cite{MPG19,MPG17,MPG17_2}. The physics-informed neural networks (PINN), as a supervised deep learning method, can solve the forward and inverse problems of partial differential equations (PDEs) with a small sample of data, and exhibits outstanding generalization capabilities. This deep learning framework suddenly became the focus, and numbers of new developments were reported based on the framework mentioned above\cite{YP19,LXG21,XSHG21,AKG20,MAG20,GLG19,DLG20}. For example, Yang et al. proposed a Bayesian physics-informed neural network (B-PINN) to solve both forward and inverse nonlinear problems described by PDEs and noisy data\cite{LXG21}. The improved PINN method presented by G.E. Karniadakis et al. which employ adaptive activation functions to accelerate the minimization process of the loss values\cite{AKG20}. M. Raissi's own team developed the idea of PINN and solved the problems related to fluid visualization of Navier-Stokes equations\cite{MAG20}. The PINN method has even been applied to fractional partial differential equations\cite{GLG19}, stochastic partial differential equations\cite{DLG20}, and so on.

Integrable systems are a class of nonlinear systems with excellent properties. The abundant exact solutions in integrable systems provide a huge sample space for the PINN method. In the integrable deep learning community, Chen's team achieved a series of results\cite{JY20,JJY21,JJC2101,WJY2105,SY}. For example, Pu et al. used an improved PINN method to recover the soliton, breathers and rogue of the nonlinear Schr\"{o}dinger (NLS) equation\cite{JJC2101}. Based on the PINN method, Peng et al. obtained the data-driven periodic rogue wave and other local wave solutions of the Chen-Lee-Liu (CLL) equation\cite{WJY2105}. Lin and Chen fully considered the advantages of the integrable systems and proposed a two-stage PINN method based on conserved quantities to solve Boussinesq-Burgers (BB) equation and Sawada-Kotera (SK) equation\cite{SY}. In addition, Wang and Yan discussed the forward and inverse problems of the defocusing nonlinear Schr\"{o}dinger equation based on the PINN method \cite{LZ21}. Dai et al. used PINN method to solve a variety of femtosecond optical soliton solutions of high order NLS equation\cite{YGYC22103}. 

Compared with the traditional method, the PINN method is robust under the $small\ data$ regime, which provides a possibility for advancing the solution of high-dimensional problems, and related researches have made attempts in high-dimensional systems\cite{GLG19,AKG20}. But for general nonlinear systems, this approach always sacrifices space cost and time cost to a large extent. The challenge of the inevitable curse of dimensionality is even worse when it comes to solving tasks that are inherently unstable. For example, the trajectory of a chaotic system has the property of exponential separation, so as the error is transmitted in the network, the predicted result will deviate from the true value. This means that chaotic systems have an extremely low tolerance for errors in the PINN. However, the rich local wave solutions in the integrable systems make it reasonable for us to discuss a small training area. Combined with the advantage of the high tolerance of the integrable systems to errors (the trajectories will not be separated exponentially), the curse of dimensionality has the possibility of being overcome. Therefore, the integrable systems may be the most suitable place for the PINN method to work, especially in high-dimensional situations.

In our work, the framework of the PINN method to solve the $(N+1)$ dimensional initial boundary value problem with $2N+1$ hyperplane boundaries is given, and many feasible model adjustments are considered. And this framework is applied to solve the most classic high-dimensional integrable equations such as: $(2+1)$-dimensional Kadomtsev–Petviashvili (KP) equation, (3+1)-dimensional reduced Kadomtsev–Petviashvili equation. More specifically, in section \ref{PINN}, the framework of the PINN method under the high-dimensional system is given; in section \ref{KP_2+1_sec}, the dynamic behaviors of the single soliton, two solitons, breathers and lump of the $(2+1)$-dimensional KP equation are reproduced based on the PINN method; in section \ref{KP_3+1_sec}, the PINN method is applied to solve the interaction solution (resonance rogue) of the $(3+1)$-dimensional reduced KP equation. Finally, conclusion and expectation are given in the last section.
\section{The PINN method in high-dimensional space}\label{PINN}
The general form of a $(N+1)$-dimensional nonlinear evolution equation in real space is expressed as
\begin{equation}\label{PINN_eq1}
	u_t+\mathcal{N}[u]=0,\ \textbf{x}\in \Omega,\ t\in [-T,T],
\end{equation}
where $u=u(\textbf{x},t)$ is the real-valued solution of the equation, $\Omega$ is a subset of $\mathbb{R}^N$, therefore $\textbf{x}$ is actually an $N$-dimensional vector recorded as $\textbf{x}=(x_1,x_2,\cdots,x_N)$, and $\mathcal{N}[\cdot]$ is a nonlinear differential operator with respect to $\textbf{x}$. Thus $\mathcal{N}[u]$ is a nonlinear functional and written as $\mathcal{N}_u(u,u_{x_1},...,u_{x_N},u_{x_1x_1},...u_{x_Nx_N},...)$, which generally contains high-order dispersion terms and nonlinear terms. In this section, we will propose a framework for the application of the PINN method in high-dimensional space.
\subsection{Neural network structure and initialization}
\quad

Consider building a feedforward neural network with a depth of $D$, which includes an input layer ($0^{th}$ layers), an output layer ($D^{th}$ layers), and $D-1$ hidden layers. The input layer contains $N+1$ nodes, and the output layer has 1 node, which depends on the dimensions of the domain and the range in equation \eqref{PINN_eq1}. Without loss of generality, assume that the number of nodes in the $d^{th}$ layer network is $N_d$. Regarding the number of network parameters, the increase in dimensionality only affects the input layer, so even considering the possibility of curse of dimensionality, the fully connected network is still reliable. The $D$-layer feedforward neural network with fully-connected structure is shown as
\begin{figure}[htbp]
\centering
\includegraphics[width=14cm,height=7cm]{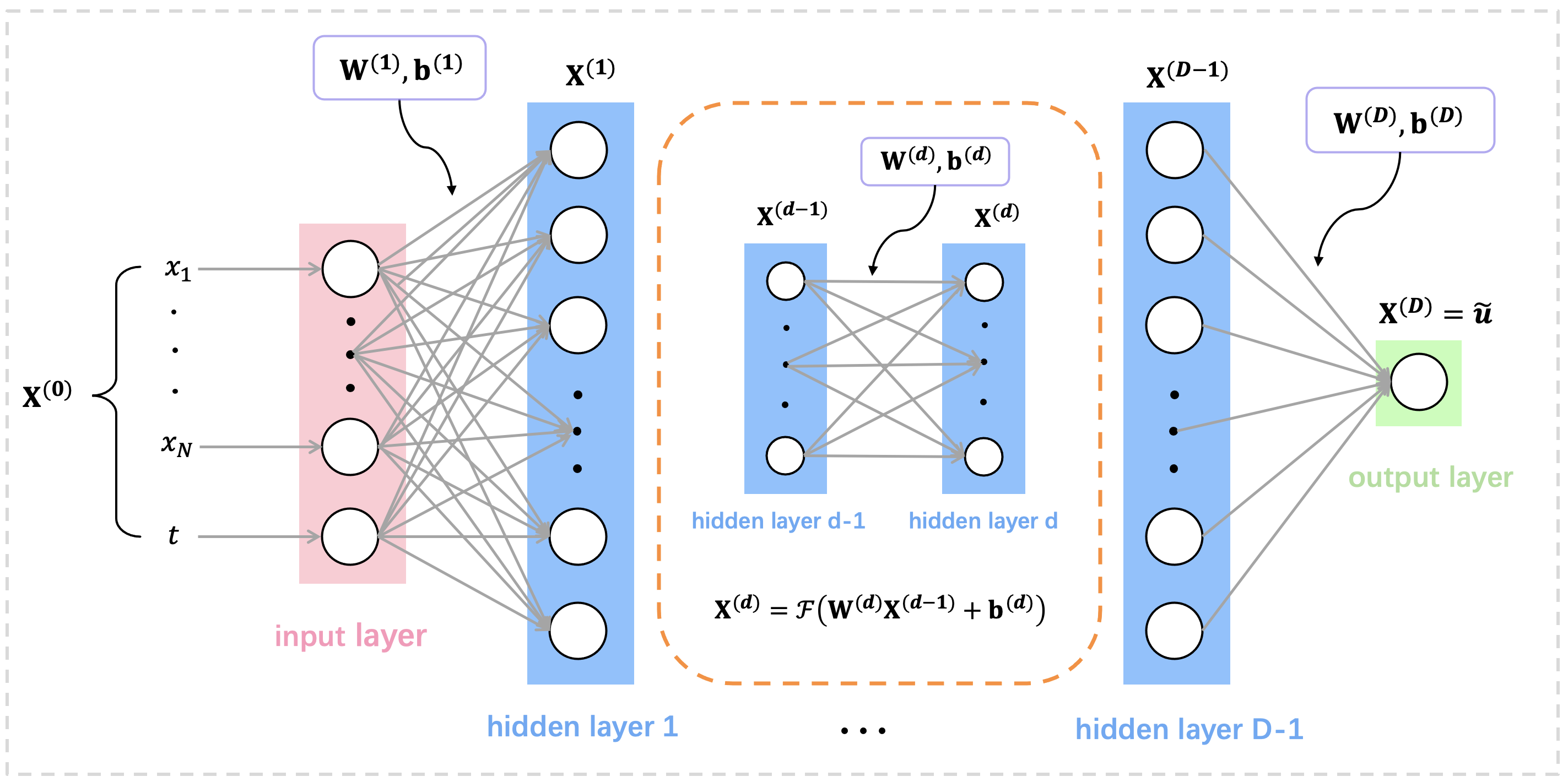}
\caption{(Online version in colour) The $D$-layer feedforward neural network with fully connected structure.}
\label{Network}
\end{figure}

The $i^{th}$ network receives the $N_i$-dimensional vector input of the previous network, and outputs the $N_{i+1}$-dimensional vector to the next network, which is clear in the network structure diagram Fig.\ref{Network}. Of course, for problems with high complexity, it is a wise choice to replace a fully-connected network with a partially connected network. In order to connect the vectors of different dimensions between the layers of the network, it is necessary to introduce a nonlinear transformation, written as
\begin{equation}\label{tf_1}
	\textbf{X}^{(d)}=\mathcal{T}_{d}(\textbf{X}^{(d-1)})=\mathcal{F}(\textbf{W}^{(d)}\textbf{X}^{(d-1)}+\textbf{b}^{(d)}),
\end{equation}
where $\textbf{W}^{(d)}\in \mathbb{R}^{N_{d} \times N_{d-1}}$ and $\textbf{b}^{(d)}\in \mathbb{R}^{N_{d}}$ represent the weight matrix and bias vector of the $d^{th}$ layer network, respectively. And $\textbf{X}^{(d)}$ is the state vector of the $d^{th}$ layer node, especially $\textbf{X}^{(0)}=\textbf{x}\oplus t=(x_1,x_2,...,x_N,t)$, and $\mathcal{F}$ is a nonlinear activation function, which is usually selected as $Sigmoid$ function, $Tanh$ function or $Relu$ function, etc. However, $\mathcal{T}_d$ is a nonlinear transformation composed of a nonlinear activation function $\mathcal{F}$ and an affine transformation, and the formula \eqref{tf_1} tells us that our model fixes the activation function of each layer. The weights and biases of the network are stored in tensor-type variables in tensorflow\cite{S17}, so they need to be initialized. For bias terms, they are initialized to tensors with an initial value of zero, but the initialization of weight terms satisfys the conditions:
\begin{equation}\label{Xavier_1}
	\textbf{W}^{(d)}\sim N(0,\sigma_d^2),\ \sigma_d^2=\frac{2}{N_d+N_{d+1}},\ d=1,2,...,D,
\end{equation}
which is the $Xavier$ initialization method\cite{A19} that can maintain state variance and gradient variance. And in fact, \eqref{Xavier_1} is a truncated normal distribution, which means that the initial value outside of two standard deviations will be discarded. In addition, methods such as $Kaiming$ initialization\cite{KXSJ16} and $Fixup$ initialization\cite{HYT19} may also be used.
\subsection{Initial boundary value condition}\label{IB_sec}
\quad

It is well known that the solution of universal equations (partial differential equations without definite solution conditions) is uncertain, but for the current deep learning technology, only one fixed solution without parameters can be learned. So the key is to establish a problem with initial boundary value. Consider the $Dirichlet\ boundary$ conditions of equation \eqref{PINN_eq1}:
\begin{equation}\label{I_b_condition}
	\begin{split}
	   \begin{cases}
		u(\textbf{x},-T)=u_0(\textbf{x},-T),\ \forall \textbf{x}\in \Omega,\\
		u(\textbf{x},t)=u_0(\textbf{x},t),\ \forall \textbf{x} \in \partial \Omega,\ t\in[-T,T],
	   \end{cases}
	\end{split}
\end{equation}
where the solutions $u_0(\textbf{x},t)$ of equation \eqref{PINN_eq1} is our learning goal, and the two formulas in \eqref{I_b_condition} represent the initial value condition and the boundary condition respectively. Of course, other boundary conditions such as $Neumann \ Boundary$ and $Periodic \ Boundary $ are also feasible at the theoretical level, and even have better performance.

\begin{figure}[htbp]
\centering
\includegraphics[width=15.2cm,height=6cm]{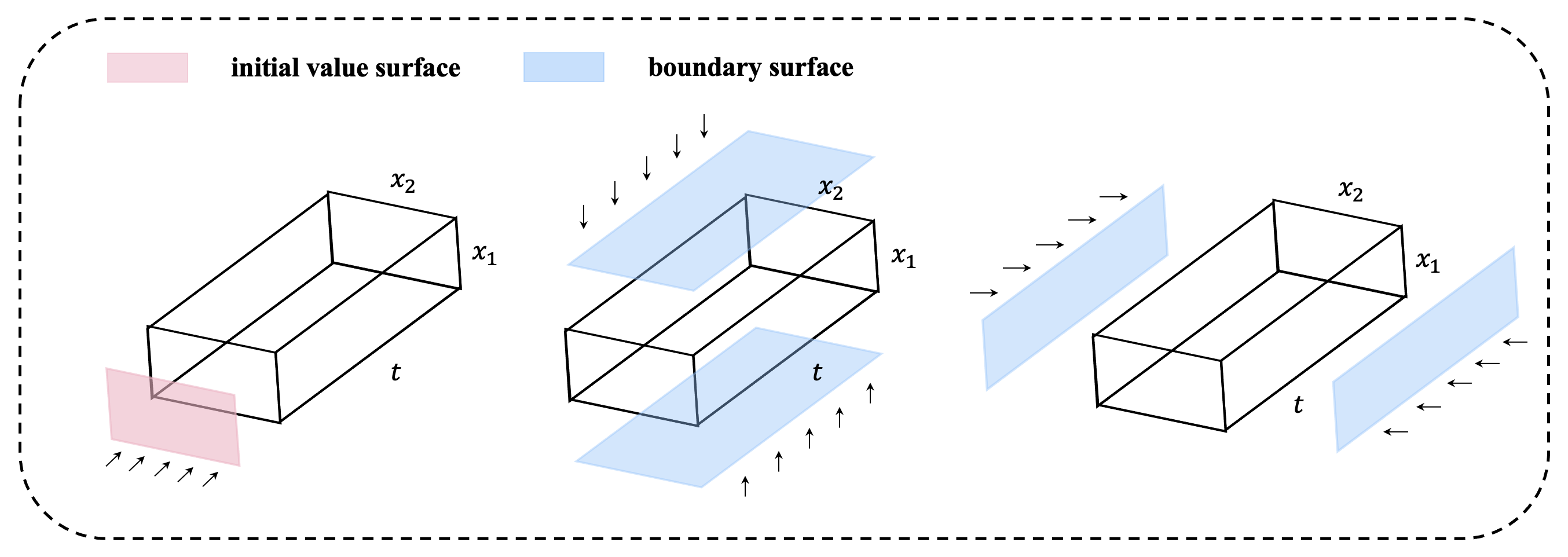}
\caption{(Online version in colour) Initial value and boundary surfaces in $(2+1)$-dimension.}
\label{IB_fig}
\end{figure}

In order to facilitate the acquisition of initial value and boundary value data for the neural network to learn, only a simple domain shape is discussed in the paper. Consider the domain $\Omega$ is an N-dimensional closed rectangular as:
\begin{equation}
	\Omega=[x_1^{l},x_1^{u}]\times[x_2^{l},x_2^{u}]\times\cdots\times[x_N^{l},x_N^{u}],
\end{equation}
Where $x_i^{u}$ and $x_i^{l}$ are defined as the upper and lower bounds of $\Omega$ in the dimension of $x_i$. Therefore, the boundary of the domain $\Omega$ is a regular $N-1$ dimensional closed rectangle, combined with the initial value conditions in \eqref{I_b_condition}, then the data points needed for deep learning will come from the following $2N+1$ hyperplanes:
\begin{equation}
H_i^l=\left\{{\textbf{X}\in \mathcal{O}\ |\ x_i=x_i^l}\right\},\ 
     H_i^u=\left\{{\textbf{X}\in \mathcal{O}\ |\ x_i=x_i^u}\right\},\ 1\le i\le N,
\end{equation}
\begin{equation}\label{H_0}
	H_0=\left\{{\textbf{X}\in \mathcal{O}\ |\ t=-T}\right\},\ \mathcal{H}=H_0 \cup \bigcup_{i=1}^{N} (H_i^l\cup H_i^u),
\end{equation}
where $\mathcal{O}=\Omega \times [-T,T]$, $\textbf{X}=(x_1,x_2,...,x_N,t)$,\ and $\mathcal{H}$ represents the union of all $2N+1$ hyperplanes. Previous work has tried to use the PINN method for (1+1)-dimensional equations, so only two boundary lines and one initial value line need to be considered. However, the work of this paper is to apply the PINN method to the (N+1)-dimensional equation, which is the first time it is proposed in the field of integrable deep learning. This means that the amount of data for the initial boundary value will increase exponentially. For example, for a $(2+1)$-dimensional system, there will be $4$ boundary surfaces and 1 initial value surface. Fig. \ref{IB_fig} displays the initial value surface and boundary surfaces in the case of $N=2$.

Deep learning requires discrete data points to support the network optimization process, as a result, the data points of the continuous equation $\eqref{PINN_eq1}$ must be discretized. First, divide the $i^{th}$ dimension of the spatial domain $\Omega$ into $n_i$ equal parts, and then divide the time domain $[-T,T]$ into $n_t$ equal parts to obtain an $N+1$ dimensional grid rectangle. We define these $(n_1+1)\cdots(n_N+1)(n_t+1)$ points as the $2$-$type$ points, and their intersection with the set $\mathcal{H}$ is additionally specified as the $1$-$type$ points. In addition, the $Latin\  hypercube \ sampling$ (LHS) method \cite{M87} is used to randomly select $n_f$ points in $\mathcal{O}$, which are defined as the $3$-$type$ points or internal collocation points. The training of the neural network and the calculation of the generalization error of the model are inseparable from these three types of points, and the specific details are discussed in the next subsection.
\subsection{Loss function, model training and generalization}\label{loss_sec}
\quad

The loss function is the key to network optimization. However, the loss function of PINN method is composed of initial-boundary condition constraints and physical equation constraints, which correspond to the $1$-$type$ points and the $3$-$type$ points, respectively. The $1$-$type$ points are defined as $\left\{{\textbf{X}_b^{i},u_b^{i}}\right\}_{i=1}^{n_b}$, and the $3$-$type$ points are defined as $\left\{{\textbf{X}_f^{i}}\right\}_{i=1}^{n_f}$, where $\textbf{X}_b^i=\textbf{x}_b^i\oplus t_b^i=(x_{b,1}^i,...,x_{b,N}^i,t_{b}^i)$, $\textbf{X}_f^i=\textbf{x}_f^i\oplus t_f^i=(x_{f,1}^i,...,x_{f,N}^i,t_{f}^i)$, and $u_b^i=u_0(\textbf{x}_b^i,t_b^i)$. Next, define a function $f(\tilde{\textbf{x}},\tilde{t},u)$ derived from the left side of the equation \eqref{PINN_eq1}, i.e.
\begin{equation}
	f(\tilde{\textbf{x}},\tilde{t};u)=\mathcal{N}_0[u]|_{\textbf{x}=\tilde{\textbf{x}},t=\tilde{t}},\ \mathcal{N}_0[\cdot]=\mathcal{N}[\cdot]+\partial_t[\cdot],
\end{equation}
where $u=u(\textbf{x},t)$ is regarded as a function parameter, especially if $u_0$ is the solution of equation \eqref{PINN_eq1}, then obviously there is $f(\textbf{x},t;u_0)=0,\forall \textbf{x}\in \Omega,t\in [-T,T]$. However, the main idea of PINN is to find suitable parameters $\textbf{W}^{(d)}$, $\textbf{b}^{(d)},\ d=1,2,...,D$ from parameter space $\Theta$ to make the composite function $\tilde{u}(\textbf{x},t,\theta)$ close enough to the solution $u_0$ of equation \eqref{PINN_eq1}. And $\tilde{u}(\textbf{x},t,\theta)$ is actually a function represented by the neural network, specifically written as
\begin{equation}\label{u_pinn}
	\tilde{u}(\textbf{x},t,\theta)=(\mathcal{T}_{D-1}\circ \mathcal{T}_{D-2}\circ \cdots \circ \mathcal{T}_1)(\textbf{X}),
\end{equation}
where $\textbf{X}=\textbf{x}\oplus t$, $\theta=\left\{{\textbf{W}^{(d)},\textbf{b}^{(d)}}\right\}_{d=1}^{D}\in \Theta$ and “$\circ$” represents the functional composition operator. Each group of parameters $\theta$ in the parameter space $\Theta$ determines a function $\tilde{u}(\textbf{x},t,\theta)$, which is obvious in \eqref{u_pinn}. In order to better measure the gap between $u_0(\textbf{x},t)$ and $\tilde{u}(\textbf{x},t,\theta)$, a loss function composed of initial-boundary condition constraints and physical equation constraints will be constructed as
\begin{equation}\label{loss_1}
	Loss(\theta)=Loss_b(\theta)+Loss_f(\theta),
\end{equation}
where
\begin{equation}\label{loss_2}
	Loss_b(\theta)=\frac{1}{n_b}\sum_{i=1}^{n_b}|\tilde{u}(\textbf{x}_b^i,t_b^i,\theta)-u_b^i|^2,\ Loss_f(\theta)=\frac{1}{n_f}\sum_{i=1}^{n_f}|f(\textbf{x}_f^i,t_f^i;\tilde{u}(\textbf{x},t,\theta))|^2.
\end{equation}
Here, $Loss_b$ represents the initial-boundary condition constraint, which measures the data fitting condition at the $1$-$type$ points, and $Loss_f$ is the physical equation constraint, which measures whether the equation is satisfied at the $3$-$type$ points. In addition, the partial derivatives involved in the model can be easily solved in tensorflow with the help of automatic differentiation technology\cite{ABA18}. Of course, because of the high-cost nature of the data, the training of initial-boundary points usually involves only a part of the $1$-$type$ points. Obviously, the smaller the loss function is, the closer $\tilde{u}(\textbf{x},t,\theta)$ is to $u_0(\textbf{x},t)$ (it just means that some of their properties are similar at specific points). The goal of model training is to find the optimal parameter $\theta^*=\left\{{\textbf{W}^{(d),*},\textbf{b}^{(d),*}}\right\}_{d=1}^{D}$, so that the loss function can reach the global minimum (of course, it must be small enough). Here, we utilize the L-BFGS-B algorithm\cite{DJ89} to optimize loss functions, in addition to gradient descent algorithms such as $stochastic\ gradient\ descent$ (SGD)\cite{L12} and $batch\ gradient\ descent$ (BGD)\cite{S17} can also be considered. Regrettably, the global optimal solution may not always be found, and some local optimal solutions are replaced. These local optimal solutions can still be accepted by us if they have good performance in model generalization.

The $2$-$type$ points mentioned above play an important role in measuring the generalization performance of the model. They are defined as $\left\{{\textbf{X}_g^{i},u_g^i}\right\}_{i=1}^{n_g}$, where $\textbf{X}_g^i=\textbf{x}_g^i\oplus t_g^i=(x_{g,1}^i,...,x_{g,N}^i,t_{g}^i)$ and $u_g^i=u_0(\textbf{x}_g^i,t_g^i)$. The training of the model does not involve the $2$-$type$ points, which is obvious from the expressions \eqref{loss_1} and \eqref{loss_2} of the loss function. Therefore, it is reasonable to define the generalization error of the model at the $2$-$type$ points, as
\begin{equation}\label{error}
Error=\frac{\sqrt{{\sum_{i=1}^{n_g}}|\tilde{u}(\textbf{x}_g^i,t_g^i,\theta^*)-u_g^i|^2}}{\sqrt{{\sum_{i=1}^{n_g}|u_g^i|^2}}},
\end{equation}
where $\tilde{u}(\textbf{x}_g^i,t_g^i,\theta^*)$ represents the result of the generalization of the optimal solution of the model at point $(\textbf{x}_g^i,t_g^i)$, and \eqref{error} is actually an error calculation based on the $\mathbb{L}_2$ norm, which mainly measures the average level of error. Other norms can construct different error formulas, such as the error formula based on the $\mathbb{L}_{\infty}$ norm to measure the maximum error of the model. The generalization error shows the generalization ability of the model. However, deep learning is to balance model training errors and model generalization errors to avoid overfitting. Therefore, it is a good choice to add parameter regularization items to the loss function, such as $L_1$ or $L_2$ regularization\cite{FMT95}.
\begin{figure}[htbp]
\centering
\includegraphics[width=15.2cm,height=11.9cm]{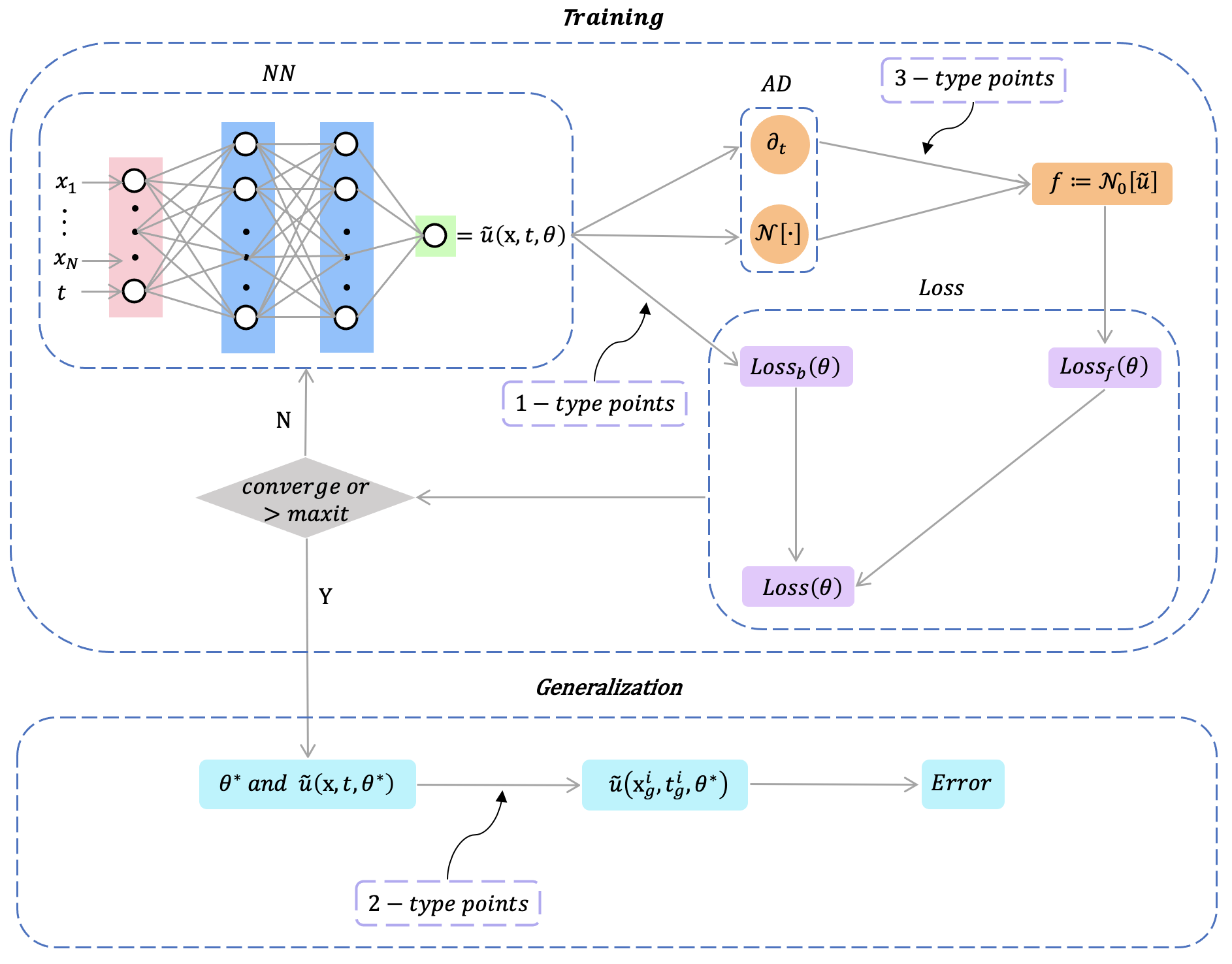}
\caption{(Online version in colour) Process sketch of the PINN method in a high-dimensional system.}
\label{sketch_PINN}
\end{figure}

Fig. \ref{sketch_PINN} is a process sketch of the PINN method in a high-dimensional system. Without loss of generality, this paper assumes that the number of nodes in the hidden layer is equal $(N_d=N_0,d=1,2,...,D)$, and selects the hyperbolic function $tanh$ as the activation function (if there are no special instructions). All codes in this article are based on Python 3.7 and tensorflow 1.15, and all numerical examples reported here are run on a DELL Precision 7920 Tower computer with 2.10 GHz 8-core Xeon Silver 4110 processor and 64 GB memory.
\section{The application of PINN method to $(2+1)$-dimensional KP equation.}\label{KP_2+1_sec}
In this section, an example of $N=2$, that is, the $(2+1)$-dimensional KP equation become the object of our discussion. The nondimensional form of the $(2+1)$-dimensional KP equation\cite{BV70} is written as
\begin{equation}\label{KP2}
	(u_t+u_{xxx}+6uu_x)_x+\sigma^2u_{yy}=0,
\end{equation}
where $\sigma^2$ is the equation parameter and $u=u(x,y,t)$. In particular, \eqref{KP2} is called the KPI when $\sigma^2=-1$, and is called the KPII when $\sigma^2=1$, and KP equation be reduced to the KdV equation when $u$ is independent of $y$. Equation \eqref{KP2} was proposed to study the evolution of small amplitude long ion acoustic waves propagating in plasma under long lateral disturbances\cite{BV70}, and was widely used in almost all fields of physics, such as shallow water waves\cite{JNH89} and nonlinear optics\cite{DYY95}. The fully integrable KP equation is a classic high-dimensional model that is often used to test new mathematical techniques.

 Consider an initial boundary value problem with Dirichlet condition for the (2+1)-dimensional KP equation:
\begin{equation}\label{I_B_problem}
	\begin{split}
	   \begin{cases}
	   (u_t+u_{xxx}+6uu_x)_x+\sigma^2u_{yy}=0,\\
		u(x,y,-T)=u_0(x,y,-T),\ \forall (x,y)\in \Omega,\\
		u(x,y,t)=u_0(x,y,t),\ \forall (x,y) \in \partial \Omega,\ t\in[-T,T],
	   \end{cases}
	\end{split}
\end{equation}
where $u_0(x,y,t)$ is usually taken as a known solution of the KP equation  and $\Omega=[x^l,x^u]\times[y^l,y^u]$. In addition, there are two special explanations. First, because of the scale transformation, the domain $[-T,T]$ of $t$ does not lose its generality. Second, the equation in \eqref{I_B_problem} is not the standard form in \eqref{PINN_eq1}, but this does not affect our results.

The following well-known transformation
\begin{equation}
	u=2(lnf)_{xx},
\end{equation}
convert the KP equation \eqref{KP2} into bilinear form
\begin{equation}\label{bilinear_1}
	(D_tD_x+D_x^4+\sigma^2D_y^2)f\cdot f=f_{tx}f-f_tf_x+ff_{xxxx}-4f_xf_{xxx}+3f_{xx}^2+\sigma^2(ff_{yy}-f_y^2)=0,
\end{equation}
where $D_t^nD_x^mf\cdot g$ is the Hirota bilinear operator\cite{H04}. It is easy to obtain the multiple solitons solution\cite{RJ76} of the KP equation from the bilinear equation \eqref{bilinear_1}, as follows:
\begin{equation}\label{n_soliton}
	u_n=2(\frac{f_{n,xx}}{f_n}-\frac{f_{n,x}^2}{f_n^2}),\  f_n=\sum_{\mu=0,1}{\rm{exp}}(\sum_{i<j}^n\mu_i\mu_jA_{ij}+\sum_{i=1}^{n}\mu_i\eta_i),
\end{equation}
where
\begin{equation}\label{n_soliton_2}
	\eta_i=k_i[x+p_iy+(-k_i^2-\sigma^2p_i^2)t]+\xi_i^{(0)},\ e^{A_{ij}}=\frac{3(k_i-k_j)^2-\sigma^2(p_i-p_j)^2}{3(k_i+k_j)^2-\sigma^2(p_i-p_i)^2},
\end{equation}
with the wave numbers $k_i,p_i$, original positions $\xi_i^{(0)}$. Here, the summation symbol $\sum_{\mu=0,1}$ is all possible combinations of $u_i=0,1(j=1,2,...)$, so it means the addition of $2^n$ terms.

Within the acceptable error, the single soliton, two solitons, and breathers of the (2+1)-dimensional KP equation based on \eqref{n_soliton} with \eqref{n_soliton_2} have been well numerically explained through the PINN method in sections \ref{1soliton-sec} to \ref{breather-sec}. However, the dynamic behavior of lump in neural networks are discussed in section \ref{lump-sec}.

\subsection{The data-driven single soliton solution}\label{1soliton-sec}
\quad

First, when the parameters of the soliton solution \eqref{n_soliton} are set as: $n=1,\sigma^2=3,k_1=3,p_1=2,\xi_1^{(0)}=0$, the single soliton solution of the KP equation \eqref{KP2} with the parameter $\sigma^2=3$ is written as
\begin{equation}
	u_1(x,y,t)=\frac{18e^{3x+6y-63t}}{(1+e^{3x+6y-63t})^2},
\end{equation}
which is an exact solution. Let $[-T,T]=[-0.2,0.2],\ \Omega=[-2,2]\times[-2,2]$, the initial boundary value problem  will be determined when $u_0(x,y,t)=u_1(x,y,t)$ is substituted into \eqref{I_B_problem}. A $11$-layer feedforward neural network with $40$ nodes in each hidden layer is constructed to find a data-driven solution to the above initial-boundary value problem. The three dimensions $x\in [-2,2]$, $y\in [-2,2]$, $t\in[-0.2,0.2]$ of the spatio-temporal data set are divided into $n_1=n_2=65$ and $n_t=33$ discrete points to obtain the $1$-$type$ and $2$-$type$ of points mentioned in section \ref{IB_sec}. The actual data training is carried out by randomly selecting $n_b=6000$ points from the $1$-$type$ points and combining with the $n_f=50000$ $3$-$type$ points selected by the LHS method. The PINN reduces the loss to $6.5699\times10^{-5}$ after $2310$ iterations in $1482.33$ seconds, and has a good generalization effect ($Error=5.7423\times10^{-4}$). 
\begin{figure}[htbp]
\centering
\includegraphics[width=17cm,height=5.5cm]{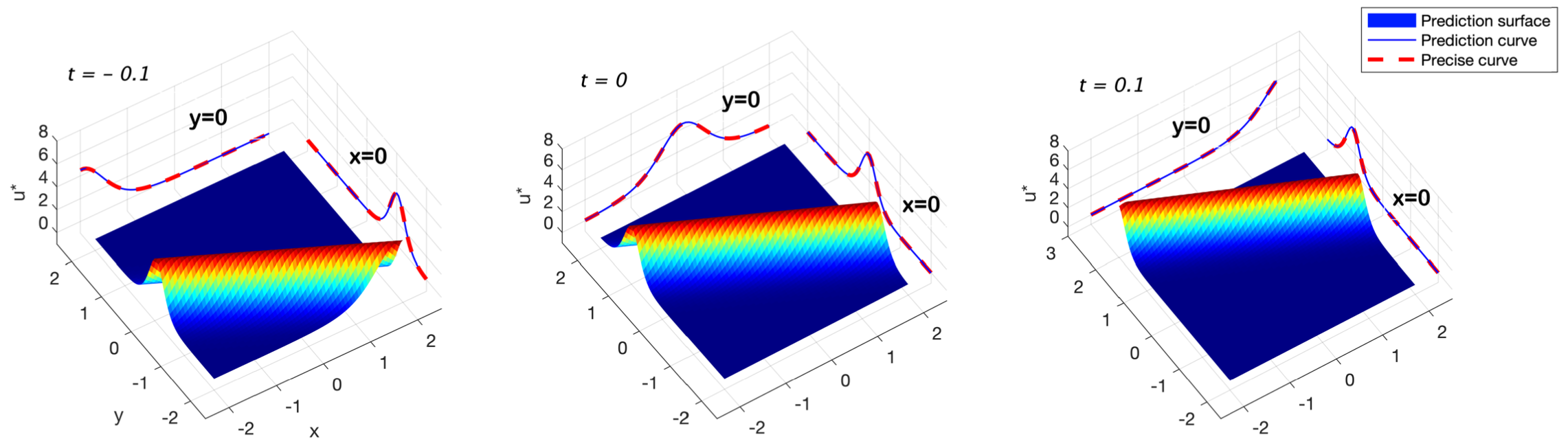}\\
$(a)$\\
\includegraphics[width=15.5cm,height=4.5cm]{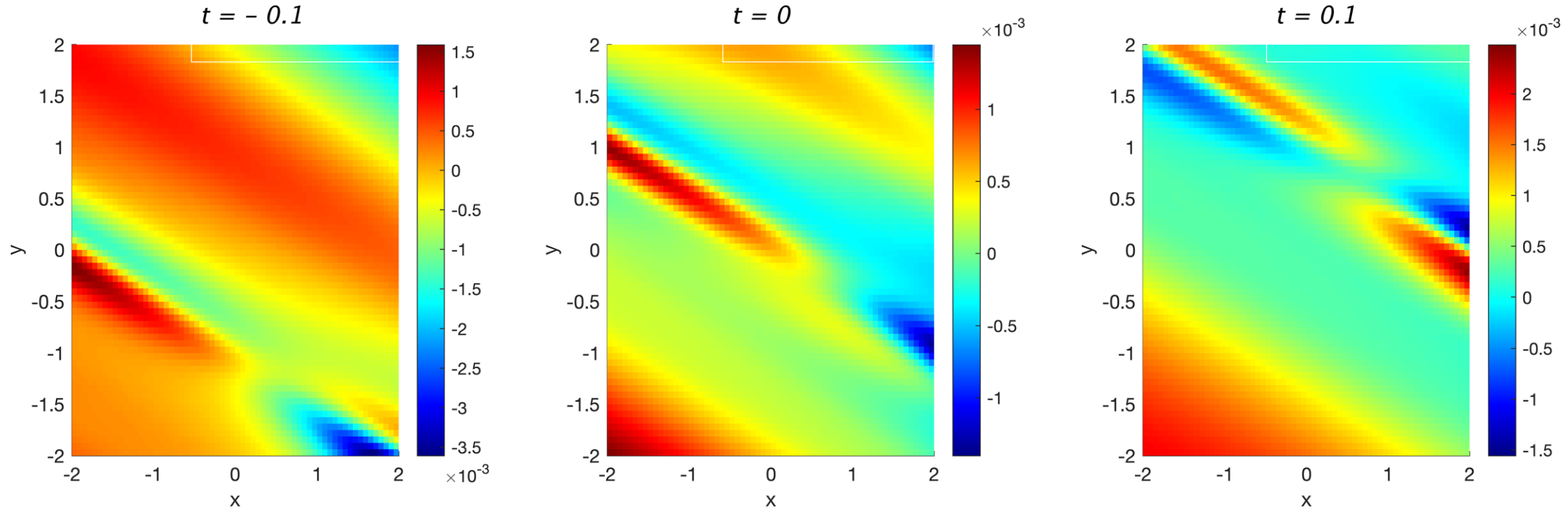}
\ \ \ \ $ $\\
$(b)$
\caption{(Online version in colour) The data-driven single soliton solution of the KP equation: (a) Three time snapshots of the three-dimensional diagram and the corresponding two-dimensional cross-sectional diagrams at $x=0$ and $y=0$ (the blue solid line and the red dashed line correspond to the predicted solution and the exact solution, respectively); (b) Density map of error (predicted solution minus exact solution).}
\label{1soliton_fig}
\end{figure}

Fig. \ref{1soliton_fig} (a) displays the time snapshot of the three-dimensional profile and two-dimensional cross-sectional view of the data-driven single soliton solution of the KP equation, which is based on the PINN method. The linear soliton evolves in the positive direction along the $x$ and $y$ axes, which is obvious in the figure. In addition, the consistency of the predicted solution and the exact solution on the cross-sectional exhibits that the model has achieved good results in the learning of single soliton solution. More specifically, the order of magnitude displayed by the color bars of the error density graph indicates that the difference between the predicted solution and the exact solution at the three moments is quite small. Fig.  \ref{1soliton_fig} (b) shows that the generalization error of the model mainly comes from the soliton rather than the background wave, which means that the learning of the neural network at large gradients needs to be improved. In general, the neural network has successfully learned the dynamic behavior of the single soliton solution, which can be derived from the above evidence and referring to the average generalization error ($Error=5.7423\times10^{-4}$).
\subsection{The data-driven two solitons solution}
\quad

Naturally, after taking the free parameter as $n=2,\sigma^2=3,k_1=-2,k_2=3,p_1=\frac{2}{5},p_2=-\frac{2}{5},\xi_1^{(0)}=-3,\xi_2^{(0)}=0$ in \eqref{n_soliton}, the two solitons solution of the KP equation with the parameter $\sigma^2=3$ is expressed as:
\begin{equation}
	u_2(x,y,t)=\frac{2(f_{2,xx}f_{2}-f_{2,x}^2)}{f_2^2},
\end{equation}
with
\begin{equation}
	f_2=1+e^{-2x-\frac{4}{5}y+\frac{224}{25}t-3}+e^{3x-\frac{6}{5}y-\frac{711}{25}t}+\frac{203}{3}e^{x-2y-\frac{487}{25}t-3},
\end{equation}

A $12$-layer feedforward neural network with 40 nodes per hidden layer will solve the initial boundary value problem \eqref{I_B_problem} with $[-T,T]=[-0.5,0.5],\ \Omega=[-4,4]\times[-2,2]$ and $u_0(x,y,t)=u_2(x,y,t)$. The discrete training data is similar to section \ref{1soliton-sec}, and the space-time dimensions are also discretely divided into $n_1=n_2=65,n_t=33$, and the corresponding number of randomly selected points of $1$-$type$ and $3$-$type$ are $n_b=6000,n_f=50000$. The PINN has also achieved great success in the two soliton learning of the KP equation, as shown in the following figure.
\begin{figure}[htpb]
\centering
\includegraphics[width=17cm,height=5.5cm]{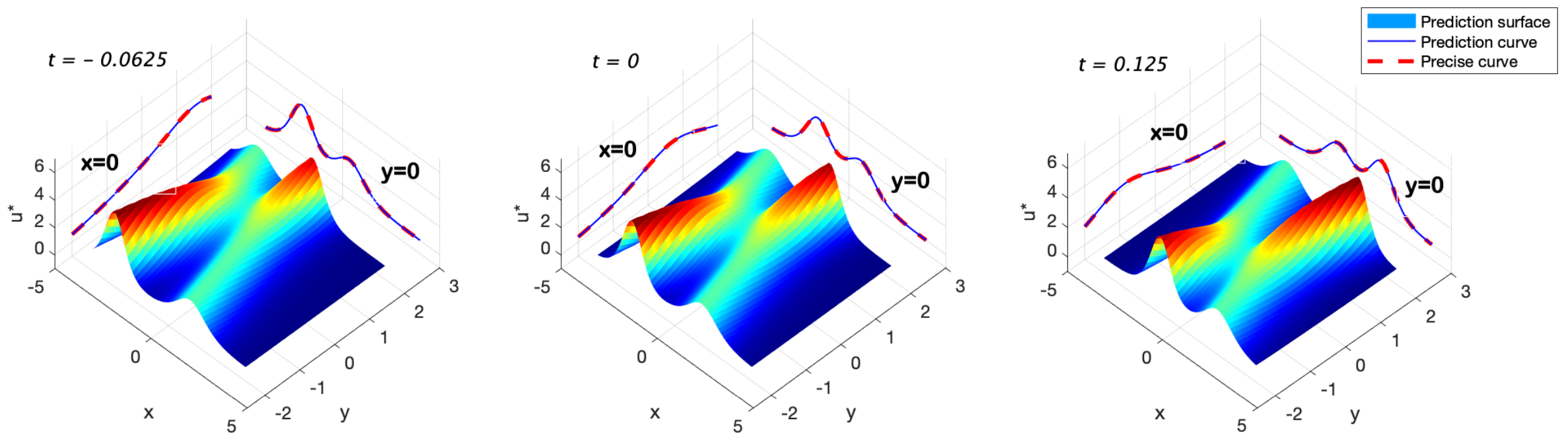}\\
$(a)$\\
\includegraphics[width=15.5cm,height=4.5cm]{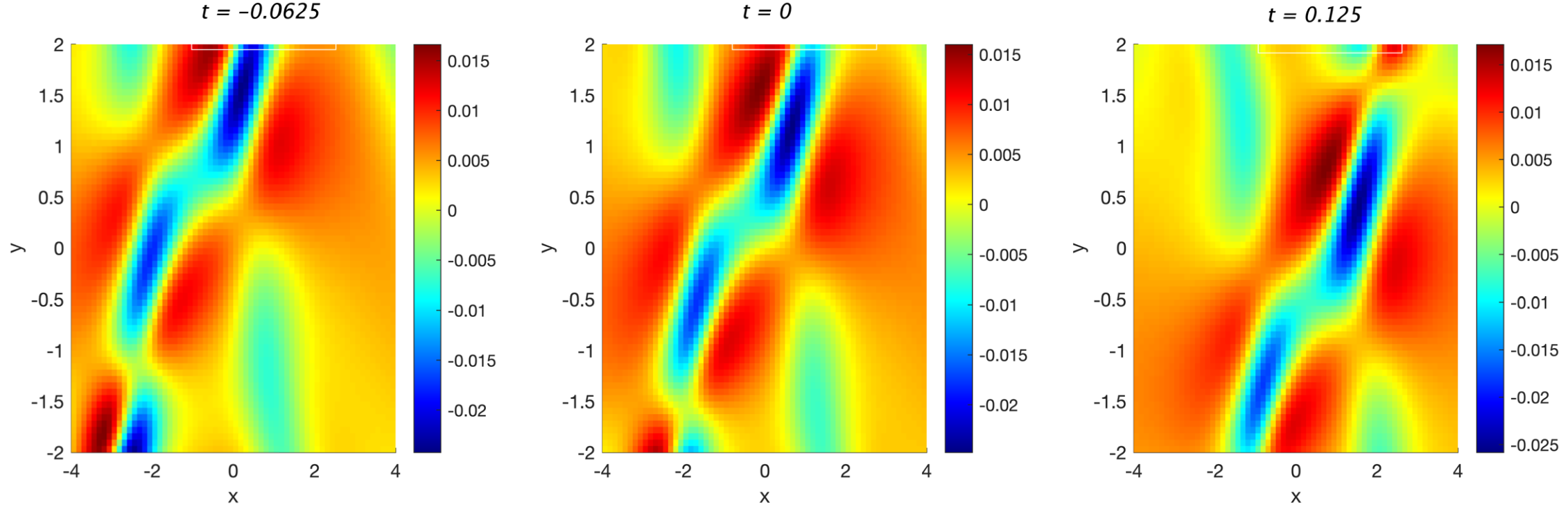}
\ \ \ \ $ $\\
$(b)$
\caption{(Online version in colour) The data-driven two solitons solution of the KP equation: (a) Three time snapshots of the three-dimensional diagram and the corresponding two-dimensional cross-sectional diagrams at $x=0$ and $y=0$ (the blue solid line and the red dashed line correspond to the predicted solution and the exact solution, respectively); (b) Density map of error (predicted solution minus exact solution).}
\label{2soliton_fig}
\end{figure}

The three-dimensional contour diagram, two-dimensional cross-section diagram and the corresponding error density diagram of the data-driven solution of the two solitons are clearly shown in Fig.\ref{2soliton_fig}. The collision of the two linear solitons in the $(2+1)$ dimension is vividly displayed. The cross-sectional views at $y=0$ at three moments fully display the process of the tall-thin soliton chasing and surpassing the short-fat soliton. From the error density map, it can found that the main error comes from the soliton with a high wave number, which also shows that a large gradient may bring a large prediction error. The error of $10^{-2}$ level is almost negligible compared to the wave height of the soliton, so the neural network once again demonstrated excellent performance. Looking back at the more specific model results, the PINN passed $11718$ iterations in $7868.21$ second, and obtained a local optimal solution so that the loss and generalization error were reduced to $7.4780\times10^{-4}$ and $4.7135\times10^{-3}$ respectively, which demonstratesed extremely low training error and excellent generalization effect.
\subsection{The data-driven breather solution}\label{breather-sec}
\quad

Parametric complexation of the two soliton solution is a common method to obtain breathers. Let $n=2$, \eqref{n_soliton} becomes a two solitons solution, then the wave number and initial position are complexed to $k1=k+i\kappa,k2=k-i\kappa,p1=p+iP,p2=p-iP,\xi_1^{(0)}=\xi+i\Xi,\xi_2^{(0)}=\xi-i\Xi$, thereby deriving the breathers of the KP equation\cite{ZSQJ07}:
\begin{equation}\label{breather_1}
	u_{breather}=2\frac{\partial^2}{\partial x^2}ln[2e^{\Omega}(cos\omega+e^{\Lambda}cosh(\Omega+\Lambda))],
\end{equation}
with 
\begin{equation}
	\Omega=kx+(kp-\kappa P)y+(2\kappa Pp\sigma^2+P^2k\sigma^2-kp^2\sigma^2+3\kappa^2k-k^3)t+\xi,
\end{equation}
\begin{equation}
	\omega=-\kappa x+(-\kappa p-Pk)y+(-\kappa^3-P^2\sigma^2\kappa+p^2\sigma^2\kappa+3k^2\kappa+2kPp\sigma^2)t-\Xi,
\end{equation}
\begin{equation}
	\Lambda=ln(\Lambda_0),\ \Lambda_0=\sqrt{\frac{P^2\sigma^2-3\kappa^2}{P^2\sigma^2+3k^2}},
\end{equation}
among them, if and only if $e^{\Lambda}>1$, \eqref{breather_1} becomes the analytical breather solution. After selecting the parameter as $\sigma^2=-3,k=\frac{1}{2},\kappa=\frac{7}{5},p=1,P=\frac{3}{5},\xi=-\frac{6}{5},\Xi=0$, a specific breather solution $u_3(x,y,t)$ is obtained. Take $[-T,T]=[-1,1],\ \Omega=[-4,4]\times[-4,4]$ and $u_0(x,y,t)=u_3(x,y,t)$ to determine the specific initial boundary value problem. After discretizing each space-time dimension into $n_1=n_2=65$ and $n_t=33$, the $LHS$ method is used to randomly select $n_b=6000$ points and $n_f=50000$ points as the training data in the $1$-$type$ and $3$-$type$ of points, respectively. In addition, because of the periodic nature of the breathers, the sine function $sin$, which also has periodicity, is more suitable as the activation function. A $10$-layer feedforward neural network with $40$ nodes per hidden layer performs $12605$ iterations within $7849.77$ seconds to reach the local optimal solution. The loss value and generalization error of this optimal solution are $1.7919\times10^{-5}$ and $1.9162\times10^{-3}$, respectively. The figure below displays that PINN has learned the breather solution well.
\begin{figure}[htpb]
\centering
\includegraphics[width=17cm,height=5.5cm]{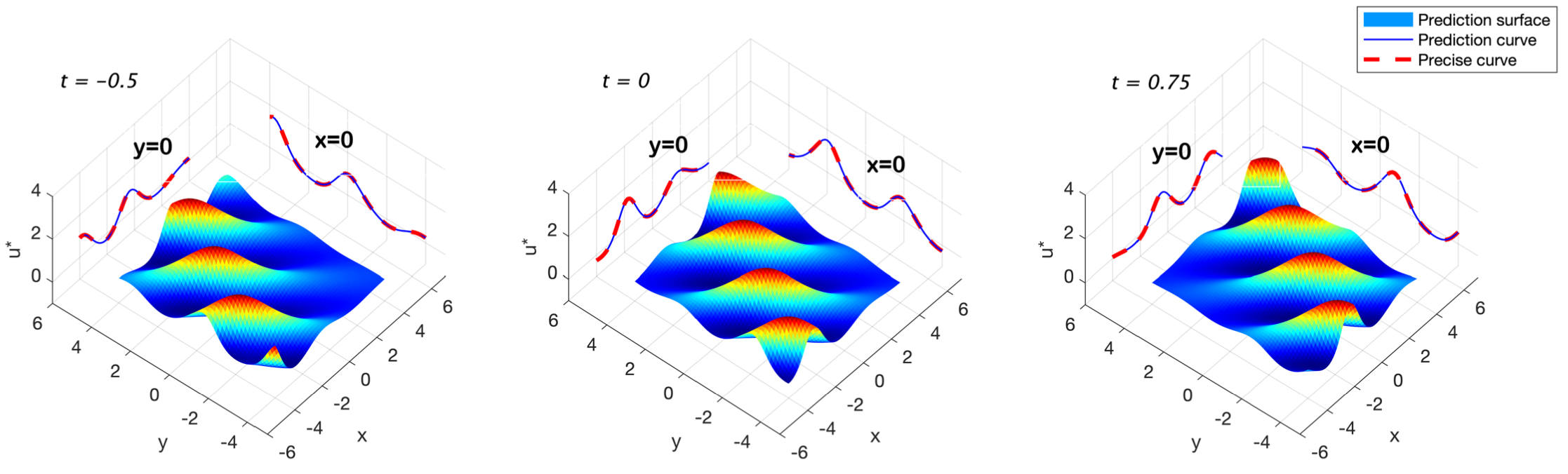}\\
$(a)$\\
\includegraphics[width=15.5cm,height=4.5cm]{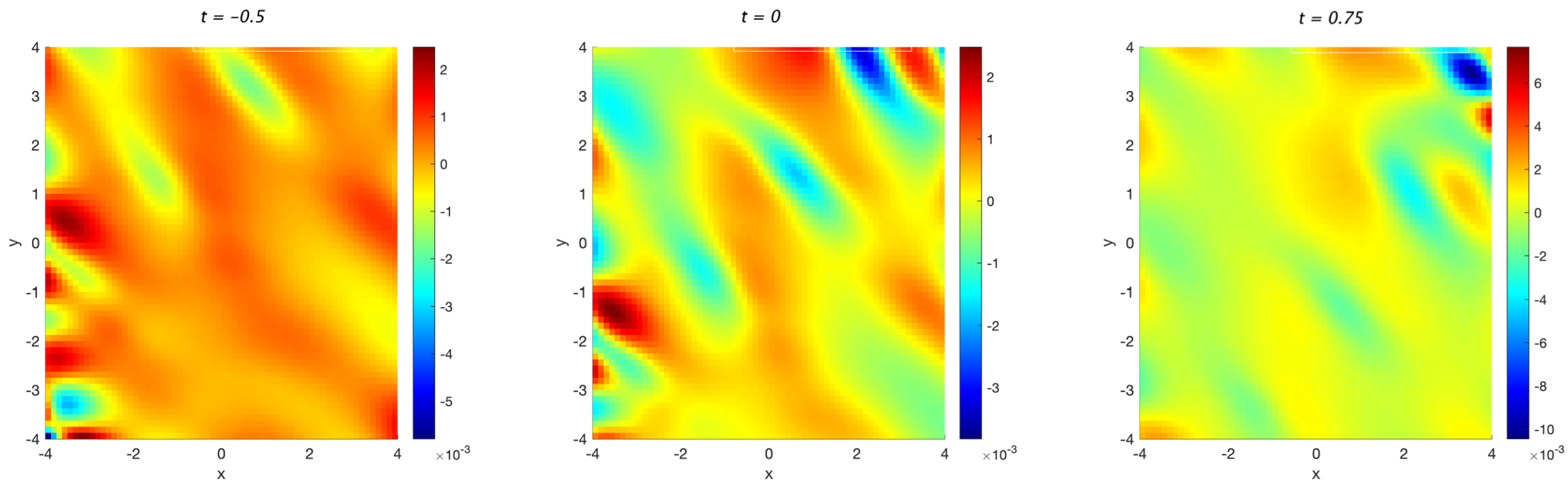}
\ \ \ \ $ $\\
$(b)$
\caption{(Online version in colour) The data-driven breather solution of the KP equation: (a) Three time snapshots of the three-dimensional diagram and the corresponding two-dimensional cross-sectional diagrams at $x=0$ and $y=0$ (the blue solid line and the red dashed line correspond to the predicted solution and the exact solution, respectively); (b) Density map of error (predicted solution minus exact solution).}
\label{breather_fig}
\end{figure}

Fig.\ref{breather_fig} displays the three-dimensional profile, the time snapshot of the two-dimensional profile and the error density map of the data-driven breather solution. The "breathing" feature of the breathers is very obvious in Fig.\ref{breather_fig}(a), and the breather wave moves to the positive direction of the $x$-axis and the negative direction of the $y$-axis as time progresses. Although the error density map is slightly blurred, it can still be found that the large gradient may be the weakness of the neural network, but the performance on the overall generalization error is impeccable. All in all, the PINN once again successfully learned the breather solution of the KP equation.
\subsection{The data-driven lump solution}\label{lump-sec}
\quad

This section will discuss the lump solution based on the PINN. Ma used the bilinear form \eqref{bilinear_1} of the KP equation to construct an analytical lump solution\cite{W15}, and $f_{lump}$ has the following form:
\begin{equation}\label{lump_1}
	f_{lump}=(a_1x+a_2y+a_3t+a4)^2+(a_5x+a_6y+a_7t+a_8)^2
	+\frac{3(a_1^2+a_5^2)^3}{(a_1a_6-a_2a_5)^2},
\end{equation}
with
\begin{equation}
	a_3=\frac{a_1a_2^2-a_1a_6^2+2a_2a_5a_6}{a_1^2+a_5^2},a_7=\frac{2a_1a_2a_6-a_2^2a_5+a_5a_6^2}{a_1^2+a_5^2},
\end{equation}
where $a_1,a_2,a_4,a_6,a_7,a_8$ are 6 free parameters, and needs to satisfy a determinant condition $a_1a_6-a_2a_5\not= 0$. And the \eqref{lump_1} is actually the solution of bilinear form with parameter $\sigma^2=-1$(KPI). Taking the free parameter as $a_1=3,a_2=4,a_4=0,a_5=-4,a_6=4,a_8=0$, the lump solution of the KPI equation is denoted as $u_4(x,y,t)$.

In the initial boundary value problem \eqref{I_B_problem}, let $[-T,T]=[-1.5,1.5],\ \Omega=[-4,4]\times[-4,4]$ and $u_0(x,y,t)=u_4(x,y,t)$. A $10$-layer neural network with $40$ nodes per hidden layer is constructed to learn the lump solution of the KP equation. The parameters involved in the discretization and random sampling of data are selected as  $n_1=n_2=65,n_t=33,n_b=5000,n_f=50000$. 
\begin{figure}[htpb]
\centering
\includegraphics[width=16.5cm,height=5.5cm]{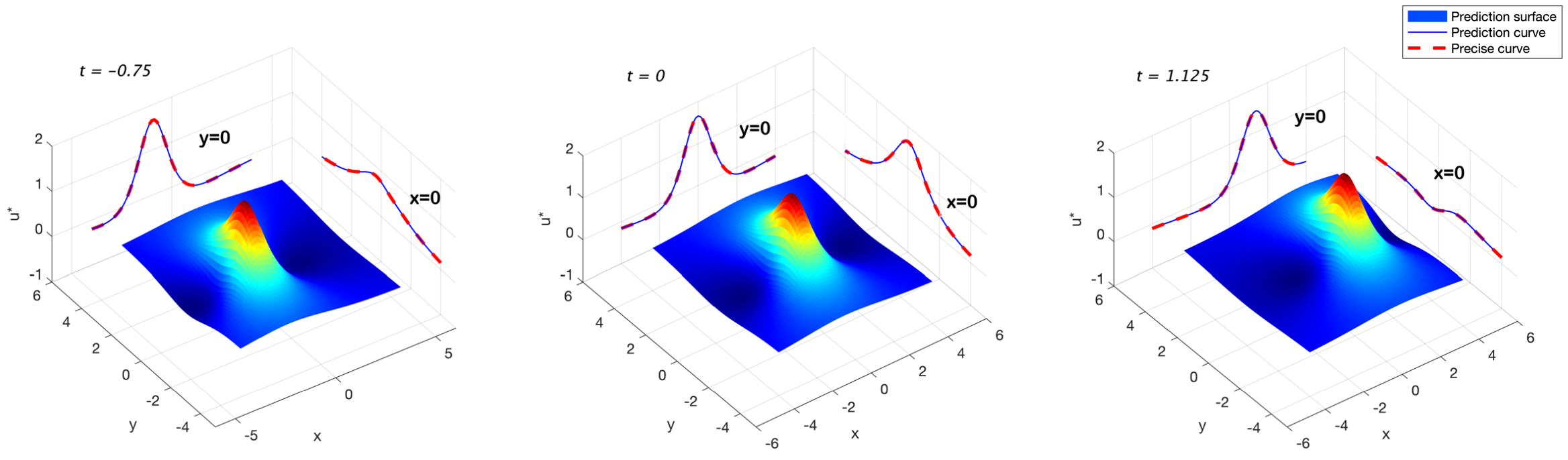}\\
$(a)$\\
\includegraphics[width=15.5cm,height=4.5cm]{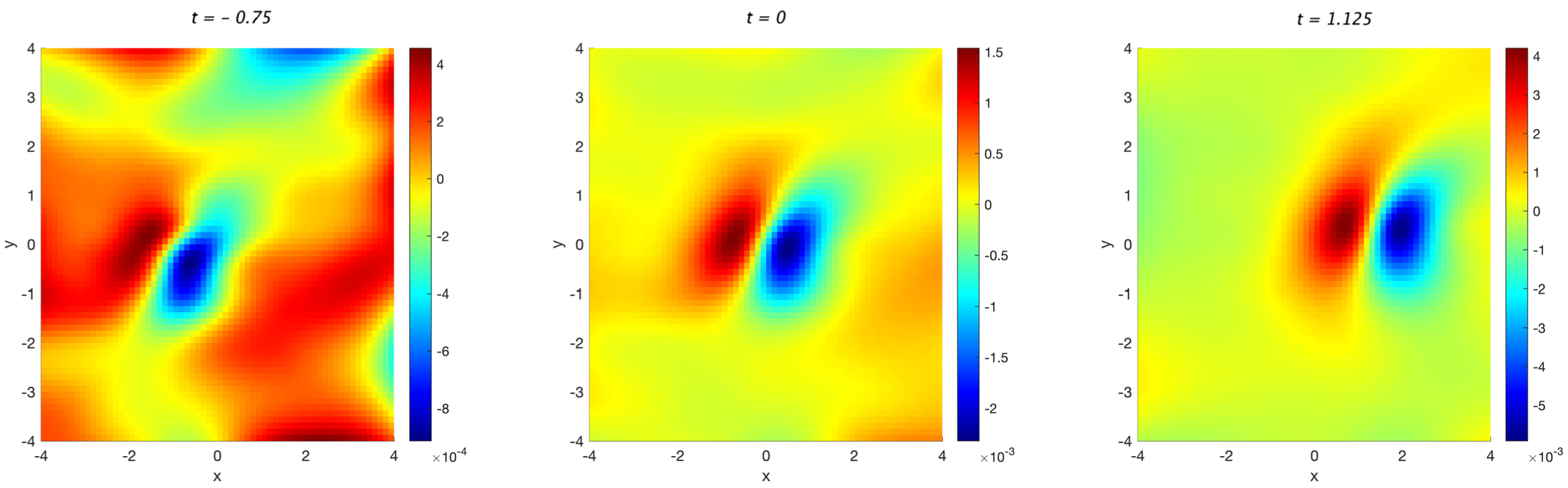}
\ \ \ \ $ $\\
$(b)$
\caption{(Online version in colour) The data-driven lump solution of the KP equation: (a) Three time snapshots of the three-dimensional diagram and the corresponding two-dimensional cross-sectional diagrams at $x=0$ and $y=0$ (the blue solid line and the red dashed line correspond to the predicted solution and the exact solution, respectively); (b) Density map of error (predicted solution minus exact solution).}
\label{lump_fig}
\end{figure}

Fig.\ref{lump_fig} displays the three-dimensional contour map, two-dimensional cross-sectional view and the corresponding error density map of the data-driven lump solution. A lump with a peak and two adjacent troughs move in the positive direction along the x-axis and y-axis, which is evident in Fig.\ref{lump_fig}(a). The two main error regions in the error density map correspond to the points near the lump extreme, which once again confirms that the generalization of large gradient regions is a challenge for PINN. The neural network undergoes $3833$ iterations in $2149.91$ seconds, and obtains a local optimal solution with a generalization error of $1.7836\times10^{-3}$. Combined with the magnitude of the error displayed by the density map, the dynamic behavior of the lump solution is well reproduced.
\section{The application of PINN method to $(3+1)$-dimensional reduced KP equation.}\label{KP_3+1_sec}
Rogue waves were first discovered in the ocean as a natural disaster. Afterwards, the rogue wave phenomenon was discovered to exist in various fields such as optics \cite{DCPB07}, Bose-Einstein condensates \cite{YVN09} and even finance \cite{Z11}, so it has become one of the most active and important research fields in experimental observation and theoretical analysis. Because of the rogue wave's characteristic of ``coming without a shadow and going without a trace", it has always been difficult to observe it. Therefore, scientists have been working to explain the mysterious dynamic behavior and formation mechanism of rogue waves through NPDEs. In this section, we use the PINN method to reproduce the dynamic behavior of a resonant rogue, which is the first discovery under the data-driven mechanism.

First of all, consider a $(3+1)$ dimensional KP equation \cite{JSMXT16}
\begin{equation}\label{KP_3+1}
	(u_t+h_1uu_x+h_2u_{xxx}+h_3u_x)_x+h_4u_{yy}+h_5u_{zz}=0,
\end{equation}
where $u=u(x,y,z,t)$ and $h_i,1\le i\le 6$ are the equation parameters. equation \eqref{KP_3+1} has a wide range of applications in plasma\cite{UAP13,ANP15}. Let the parameters such as $h_1=-1,h_2=-\frac{1}{3},h_3=1,h_4=1,h_5=-\frac{2}{3}$, and take $x=z$, equation \eqref{KP_3+1} is reduced to
\begin{equation}
	(u_t-uu_x-\frac{1}{3}u_{xxx}+\frac{1}{3}u_x)_x+u_{yy}=0,
\end{equation}
through variable transformation $u=4(lnf)_{xx}$, its bilinear equation is obtained:
\begin{equation}\label{KP_3+1_bl}
\begin{split}
   &(D_xD_t-\frac{1}{3}D_x^4+\frac{1}{3}D_x^2+D_y^2)f\cdot f=\\
   &2f_{xy}f-2f_xf_t-\frac{2}{3}f_{xxxx}f+\frac{8}{3}f_{xxx}f_{x}-2f_{xx}^2+\frac{2}{3}f_{xx}f-\frac{2}{3}f_x^2+2f_{yy}f-2f_y^2=0,
\end{split}
\end{equation}
where $D_x^nD_y^mD_t^l\ f\cdot g$ is the Hirota bilinear operator\cite{H04}. Rational solutions account for the vast majority of the known rogue wave theories. Zhang et al. received inspiration from the interaction between lump and single soliton and proposed a new nonlinear mechanism for generating rogue waves \cite{XYX18}, recently. They introduced a new combination of positive quadratic function and hyperbolic function, and obtained the resonant rogue of the $(3+1)$-dimensional reduced KP equation. This resonant rogue is actually the interaction solution between lump and resonant two solitons, based on the bilinear form \eqref{KP_3+1_bl}, $f$ is written as
\begin{equation}\label{rogue_f}
	f=g^2+h^2+\lambda {\rm cosh}(c_1x+c_2y+c_3t)+b_9,
\end{equation}
with 
\begin{equation}\label{rogue_f_2}
	g=b_1x+b_2y+b_3t+b_4,h=b_5x+b_6y+b_7t+b_8,
\end{equation}
where the lump part is expressed as quadratic functions, and the two soliton part is expressed as a hyperbolic function. Substitute \eqref{rogue_f} into equation \eqref{KP_3+1_bl} leads to:
\begin{equation}\label{rogue_f_3}
	\begin{split}
		&b_6=\frac{b_2b_5\pm b_1^2c_1\pm b_5^2c_1}{b_1},b_9=\frac{\lambda^2c_1^4+4b_1^4+8b_1^2b_5^2+4b_5^4}{4c_1^2(b_1^2+b_5^2)},c_2=\frac{c_1(b_2\pm b_5c_1)}{b_1},\\
		&c_3=\frac{c_1(b_1^2c_1^2-3b_5^2c_1^2\mp 6b_2b_5c_1-b_1^2-3b_2^2)}{3b_1^2},b_3=\frac{3b_1^2c_1^2+3b_5^2c_1^2-b_1^2-3b_2^2}{3b_1},\\
		&b_7=-\frac{3b_1^2b_5c_1^2+3b_5^3c_1^2\pm6b_1^2b_2c_1\pm6b_2b_5^2c_1+b_1^2b_5+3b_2^2b_5}{3b_1^2}.
	\end{split}
\end{equation}

Therefore, under the first type solution, the solution $u$ is
\begin{equation}
	u=\frac{4(2b_1^2+2b_5^2+\lambda {\rm{cosh}}(c_1x+c_2y+c_3t)c_1^2)}{f}-4\frac{(2b_1g+2b_5h+\lambda c_1{\rm{sinh}}(c_1x+c_2y+c_3t))^2}{f^2},
\end{equation}
where $f,g,h$ satisfy the constraints in \eqref{rogue_f}-\eqref{rogue_f_3}. Take the parameter $b_1=\frac{3}{5},b_2=-\frac{3}{5},b_4=b_5=b_8=0,\lambda=\frac{6}{5},c_1=\frac{6}{5}$ to obtain a specific interaction solution $u_5(x,y,t)$.
\subsection{The data-driven interaction solution —— resonance rogue}\label{rogue-sec}
\quad

The peaks of rogue waves usually have a larger gradient, so the learning of rogue waves is more difficult than other local waves. Next, we will discuss the performance of the PINN method on the high-dimensional rogue. First, construct an initial boundary value problem for the $(3+1)$-dimensional reduced KP equation
\begin{equation}\label{I_B_problem_2}
	\begin{split}
	   \begin{cases}
	   (u_t-uu_x-\frac{1}{3}u_{xxx}+\frac{1}{3}u_x)_x-\frac{2}{3}u_{yy}=0,\\
		u(x,y,-T)=u_0(x,y,-T),\ \forall (x,y)\in \Omega,\\
		u(x,y,t)=u_0(x,y,t),\ \forall (x,y) \in \partial \Omega,\ t\in[-T,T].
	   \end{cases}
	\end{split}
\end{equation} 
where select $[-T,T]=[-6,6],\ \Omega=[-6,6]\times[-6,6]$ and $u_0(x,y,t)=u_5(x,y,t)$. A $11$-layer neural network with $40$ nodes per hidden layer is constructed to learn the interaction solution of the $(3+1)$-dimensional reduced KP equation. The parameters involved in the discretization and random sampling of data are selected as $n_1=n_2=n_t=65,n_b=10000,n_f = 50000$. The neural network undergoes $3616$ iterations in $2368.08$ seconds, and obtains a local optimal solution with a generalization error of $2.6201\times10^{-3}$. The data-driven interaction solution and the corresponding error are shown as
\begin{figure}[htpb]
\centering
\includegraphics[width=16.5cm,height=5.5cm]{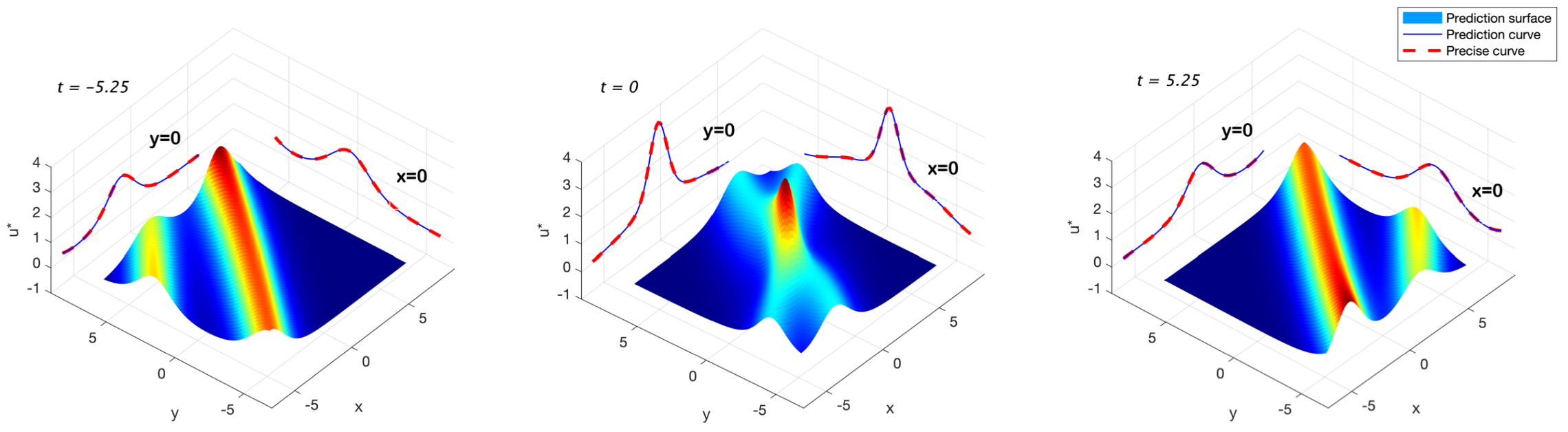}\\
$(a)$\\
\includegraphics[width=15.5cm,height=4.5cm]{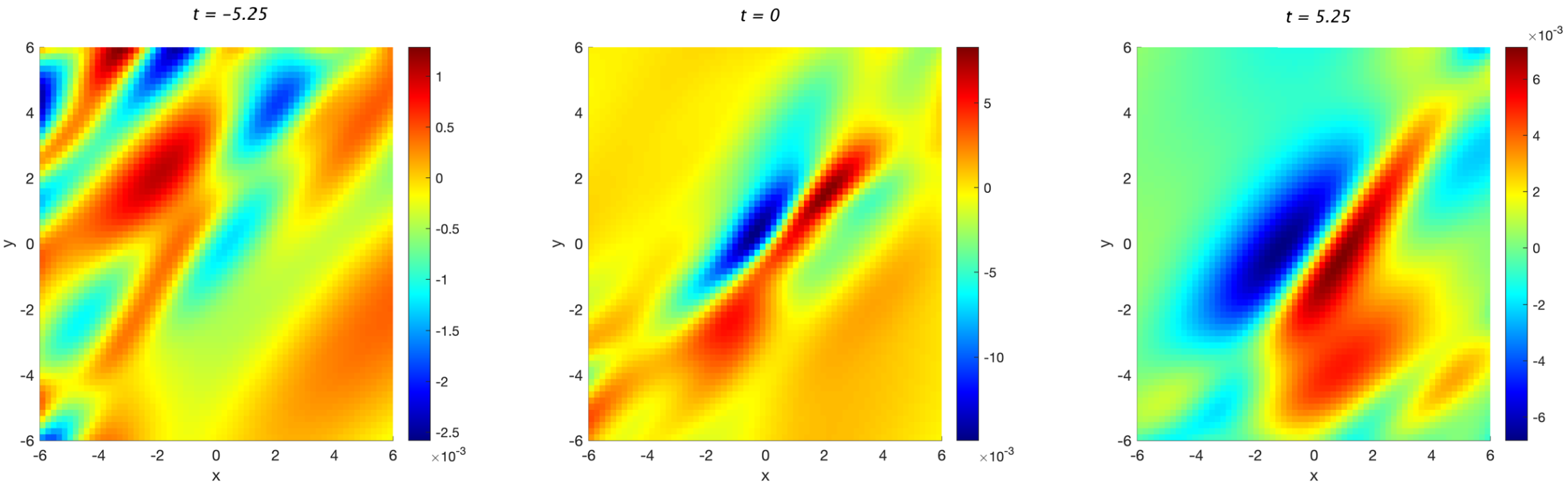}
\ \ \ \ $ $\\
$(b)$
\caption{(Online version in colour) The data-driven interaction solution of the reduced KP equation: (a) Three time snapshots of the three-dimensional diagram and the corresponding two-dimensional cross-sectional diagrams at $x=0$ and $y=0$ (the blue solid line and the red dashed line correspond to the predicted solution and the exact solution, respectively); (b) Density map of error (predicted solution minus exact solution).}
\label{rogue_fig}
\end{figure}

Fig.\ref{rogue_fig}(a) shows that the lump is hidden like a ghost soliton and only appears in the middle of the two linear solitons, that is to say, the resonance rogue appears near $x=y=t=0$. The main reason for this phenomenon is that the algebraic decay of lump is slower than the exponential decay of the resonance two solitons. The generalization error mainly comes from the larger gradient area, which is also obvious in (b). The magnitude of the error tells us that PINN successfully reproduced the dynamic behavior of the resonance rogue. This proves once again that the combination of integrable system and PINN will have unexpected results.
\section{Conclusion}
The PINN method based on physical constraints provides a possibility for solving high-dimensional problems. But the instability of the system itself will exacerbate the difficulties caused by the curse of dimensionality. As a class of NPDEs with excellent properties, integrable systems have abundant local wave solutions and high tolerance to errors (the trajectories will not be exponentially separated).

Here, the PINN method is applied to integrable systems, after fully considering the advantages of both. The framework for applying the PINN method to solve the $(N+1)$-dimensional initial boundary value problem with $2N+1$ hyperplane boundaries is presented in this paper. And at low time and space cost, the dynamic behavior of various local waves in the high-dimensional integrable system is reproduced, such as: single soliton, two solitons, breathers, lump of the $(2+1)$-dimensional KP equation, and the interaction solution (resonance rogue) of the $(3+1)$-dimensional reduced KP equation. From the error results of the model, the PINN shows outstanding generalization ability even in the high-dimensional integrable systems. This proves that the integrable systems will be suitable places for the PINN method to function, especially in the case of high-dimensional.

However, from the error density map, we discovered that the error mainly comes from the region of large gradient, which means that network optimization for the region of large gradient may be the direction of future efforts. In addition, more interesting structures in integrable systems may speed up the process of minimizing the loss function or improve the accuracy of generalization. These will become our future work.


\begin{thebibliography}{99}
\bibitem{TKRJ15}
T. Gustafsson, K.R. Rajagopal, R. Stenberg, J. Videman, Nonlinear Reynolds equation for hydrodynamic lubrication, Appl. Math. Model. 39, 5299–5309(2015).

\bibitem{DA78}
D. J. Kaup, A.C. Newell, An exact solution for a derivative nonlinear Schro\"{o}dinger equation, J. Math. Phys. 19, 798-801(1978).

\bibitem{AA14}
A.D. Polyanin, A.I. Zhurov, The functional constraints method: Application to non-linear delay reaction–diffusion equations with varying transfer coefficients, Int. J. Nonlin. Mech. 67, 267–277(2014).

\bibitem{AD98}
A.S. Parkins, D.F. Walls, The physics of trapped dilute-gas Bose-Einstein condensates, Phys. Rep. 303, 1-80(1998).

\bibitem{VM91}
V.B. Matveev, M.A. Salle, Darboux transformations and solitons, Springer-Verlag(1991).

\bibitem{MP91}
M.J. Ablowitz, P.A. Clarkson, Solitons, nonlinear evolution equations and inverse scattering, Cambridge University Press(1991).

\bibitem{CW02}
C. Rogers, W.K. Schief, B\"{a}cklund and Darboux transformations: geometry and modern applications in soliton theory, Cambridge University Press(2002).

\bibitem{H04}
R. Hirota, The direct method in soliton theory, Cambridge University Press(2004).

\bibitem{AIG12}
A. Krizhevsky, I. Sutskever, G.E. Hinton, Imagenet classification with deep convolutional neural networks, Advances in Neural Information Processing Systems, 1097–1105(2012).

\bibitem{BRJ15}
B.M. Lake, R. Salakhutdinov, J.B. Tenenbaum, Human-level concept learning through probabilistic program induction, Science, 350, 1332–1338(2015).

\bibitem{YYG15}
Y. LeCun, Y. Bengio, G. Hinton, Deep learning, Nature, 521, 436–444(2015).

\bibitem{WIO1409}
W. Zaremba, I. Sutskever, O. Vinyals, Recurrent Neural Network Regularization, arXiv:1409.2329v5.

\bibitem{MPG19}
M. Raissi, P. Perdikaris, G.E. Karniadakis, Physics-informed neural networks: A deep learning framework for solving forward and
inverse problems involving nonlinear partial differential equations, J. Comput. Phys., 378, 686-707(2019).

\bibitem{MPG17}
 M. Raissi, P. Perdikaris, G.E. Karniadakis, Inferring solutions of differential equations using noisy multi-fidelity data, J. Comput. Phys., 335, 736–746(2017).

\bibitem{MPG17_2}
M. Raissi, P. Perdikaris, G.E. Karniadakis, Machine learning of linear differential equations using Gaussian processes, J. Comput. Phys., 348, 683–693(2017).

\bibitem{YP19}
Y.B. Yang, P. Perdikaris, Adversarial uncertainty quantification in physics-informed neural networks, J. Comput. Phys., 394, 136-152(2019).

\bibitem{LXG21}
L. Yang, X.H. Meng, G.E. Karniadakis, B-PINNs: Bayesian physics-informed neural networks for forward and inverse PDE problems with noisy data, J. Comput. Phys., 425, 109913(2021).

\bibitem{XSHG21}
X.W. Jin, S.Z. Cai, H. Li, G.E. Karniadakis, NSFnets (Navier-Stokes flow nets): Physics-informed neural networks for the incompressible Navier-Stokes equations, J. Comput. Phys., 426, 109951(2021). 


\bibitem{AKG20}
A.D. Jagtap, K. Kawaguchi, G.E. Karniadakis, Adaptive activation functions accelerate convergence in deep and physics-informed neural networks, J. Comput. Phys., 404, 109136(2020).

\bibitem{MAG20}
M. Raissi, A. Yazdani, G.E. Karniadakis, Hidden fluid mechanics: Learning velocity and pressure fields from flow visualizations, Science, 367, 1026-1030(2020).

\bibitem{GLG19}
G. Pang, L. Lu, G. E. Karniadakis. fPINNs: Fractional physics-informed neural networks. SIAM Journal on Scientific Computing, 41(4), A2603–A2626(2019).

\bibitem{DLG20}
D. Zhang, L. Guo, G.E. Karniadakis. Learning in modal space: Solving time-dependent stochastic pdes using physics-informed neural networks. SIAM Journal on Scientific Computing, 42(2), A639–A665(2020).

\bibitem{JY20}
J. Li, Y. Chen, Solving second-order nonlinear evolution partial differential equations using deep learning, Commun. Theor. Phys., 72, 105005(2020). 

\bibitem{JJY21}
J.C. Pu, J. Li, Y. Chen, Soliton, breather and rogue wave solutions for solving the nonlinear Schr\"{o}dinger equation using a deep learning method with physical constraints. Chin. Phys. B, 30, 060202(2021).

\bibitem{JJC2101}
J.C. Pu, J. Li, Y. Chen, Solving localized wave solutions of the derivative nonlinear Schr\"{o}dinger equation using an improved PINN method, arXiv:2101.08593v1.


\bibitem{WJY2105}
W.Q. Peng, J.C. Pu, Y. Chen, PINN deep learning for the Chen-Lee-Liu equation: rogue wave on the periodic background, arXiv:2105.13026v1.


\bibitem{SY}
S.N. Lin, Y. Chen, A two-stage physics-informed neural network method based on conserved quantities and applications in localized wave solutions, arXiv:2107.01009.

\bibitem{LZ21}
L. Wang, Z.Y. Yan, Data-driven rogue waves and parameter discovery in the defocusing nonlinear Schr\"{o}dinger equation with a potential using the PINN deep learning, Phys. Lett. A, 404, 127408(2021).
\bibitem{YGYC22103}
Y. Fang, G.Z. Wu, Y.Y. Wang, C.Q. Dai, Data-driven femtosecond optical soliton excitations and parameters discovery of the high-order NLSE using the PINN, arXiv:2103.16297.

\bibitem{S17}
S. Pattanayak, Pro deep learning with tensorflow, Springer(2017).

\bibitem{A19}
A. Ghatak, Initialization of Network Parameters, Springer-Verlag(2019).

\bibitem{KXSJ16}
K.M. He, X.Y. Zhang, S.Q. Ren, J. Sun, Delving Deep into Rectifiers:
Surpassing Human-Level Performance on ImageNet Classification, ICCV, 1026-1034(2016).

\bibitem{HYT19}
H.Y. Zhang, Y.N. Dauphin, T.Y. Ma, Fixup initialization: Residual learning without normalization, ICLR(2019).

\bibitem{M87}
M. Stein, Large sample properties of simulations using Latin hypercube sampling, Technometrics, 29, 143-151(1987).

\bibitem{ABA18}
A.G. Baydin, B.A. Pearlmutter, A.A. Radul, J.M. Siskind, Automatic differentiation in machine learning: a Survey, J. Mach. Learning Research, 18(153), 1-43(2018).

\bibitem{DJ89}
D.C. Liu, J. Nocedal, On the limited memory BFGS method for large scale optimization, Math. Program, 45, 503-528(1989).

\bibitem{L12}
L. Bottou, Neural networks: tricks of the trade, Springer(2012).

\bibitem{S17}
S. Ruder, An overview of gradient descent optimization, arXiv:1609.04747.

\bibitem{FMT95}
F. Girosi, M. Jones, T. Poggio, Regularization theory and neural networks architectures, Neural Compu., 7(2), 219-269(1995).

\bibitem{BV70}
B.B. Kadomtsev, V.I. Petviashvili, On the stability of solitary waves in weakly dispersing media, Sov. Phys. Dokl., 15, 539(1970).

\bibitem{JNH89}
J. Hammack, N. Scheffner, H. Segur, Two-dimensional periodic waves in shallow water, J. Fluid Mech., 209, 567–589(1989).

\bibitem{DYY95}
D.E. Pelinovsky, Y.A. Stepanyants, Y.S. Kivshar, Self-focusing of plane dark solitons in nonlinear defocusing media, Phys. Rev. E, 51, 5016–5026(1995).

\bibitem{RJ76}
R. Hirota, J. Satsuma, N-Soliton of the K-dV equation with loss and Nonuniformity terms, J. Phys. Soc. Jpn., 40, 286(1976).

\bibitem{ZSQJ07}
Z.D. Dai, S.L. Li, Q.Y. Dai, J. Huang, Singular periodic soliton solutions and resonance for the Kadomtsev–Petviashvili equation, Chaos, Solitons and Fractals, 34, 1148–1153(2007).

\bibitem{W15}
W.X. Ma, Lump solutions to the Kadomtsev–Petviashvili equation, Phys. Lett. A, 379, 1975-1978(2015).

\bibitem{DCPB07}
D.R. Solli, C.Ropers, P.Koonath, B.Jalali, Optical rogue waves, Nature, 450, 1054-1058(2007).

\bibitem{YVN09}
Y.V. Bludov, V.V. Konotop, N. Akhmediev, Matter rogue waves, Phys. Rev. A, 80, 033610(2009).

\bibitem{Z11}
Z.Y. Yan, Vector financial rogue waves, Phys. Lett. A, 375, 4274-4279(2011).

\bibitem{JSMXT16}
J.M. Tu, S.F. Tian, M.J. Xu, X.Q. Song, T.T. Zhang, B\"{a}cklund transformation, infinite conservation laws and periodic wave solutions of a generalized (3+1)-dimensional nonlinear wave in liquid with gas bubbles, Nonlinear Dyn., 83, 1199–1215(2016).

\bibitem{UAP13}
U.K. Samanta, A. Saha, P. Chatterjee, Bifurcations of dust ion acoustic travelling waves in a magnetized dusty plasma with a q-nonextensive electron velocity distribution, Phys. Plasmas, 20, 022111(2013).

\bibitem{ANP15}
A. Saha, N. Pal, P. Chatterjee, Bifurcation and quasiperiodic behaviors of ion acoustic waves in magnetoplasmas with nonthermal electrons featuring tsallis distribution, Braz. J. Phys., 45, 325–333(2015).

\bibitem{XYX18}
X.E. Zhang, Y. Chen, X.Y. Tang, Rogue wave and a pair of resonance stripe solitons to KP equation, Comput. Math. Appl., 76, 1938-1949(2018).


\end{thebibliography}
\end{document}